\documentclass[11pt,letterpaper,twocolumn]{article}
\usepackage[utf8]{inputenc}
\usepackage[margin=1.5cm]{geometry}
\usepackage[T1]{fontenc}
\usepackage{mathptmx}
\usepackage{fancyhdr} 
\usepackage[pdftex]{graphicx}
\graphicspath{ {./Images/} }
\usepackage{caption}
\usepackage{hyperref}
\usepackage{gensymb}
\usepackage{multirow}

\usepackage{graphics}
\usepackage{array}
\usepackage{subcaption}
\usepackage{xcolor}
\small
\usepackage{float}
\usepackage{algorithm}
\usepackage{amsmath}
\usepackage{setspace}
\usepackage{textcomp}
\usepackage[]{siunitx}
\usepackage[left]{lineno}
\usepackage{etoolbox}
\usepackage{dblfloatfix}
\hypersetup{
    colorlinks=true,      
    linkcolor=blue,       
    citecolor=blue,       
    filecolor=blue,       
    urlcolor=blue         
}
\usepackage[
backend=biber,
style=numeric-comp,
sorting=none
]{biblatex}
\AtBeginBibliography{\footnotesize}
\begin{document}
\sloppy

\twocolumn
[
\begin{@twocolumnfalse}
\centering
\centering
{\Large\textbf{Optomicrofluidic measurement of particle-encapsulated droplet system} }

\centering

\vspace*{4mm} \noindent
{\large\textbf{
{Kanimozhi Kumaresan}$^{\textbf{1,2}}$, Thaipally Sujith$^{\textbf{2}}$, Anil Prabhakar$^{\textbf{1}}$, Ashis Kumar Sen $^{\textbf{2}}$ }}

\vspace*{1mm} \noindent
\fontsize{9pt}{16pt}\selectfont\textit{$^1$Department of Electrical Engineering,Indian Institute of Technology Madras,Chennai,India.}\\  

\fontsize{9pt}{16pt}\selectfont\textit{$^2$Department of Mechanical Engineering,Indian Institute of Technology Madras,Chennai,India.}\\  

\fontsize{10pt}{16pt}\selectfont\text{Corresponding author: Ashis Kumar Sen (ashis@iitm.ac.in) or Anil Prabhakar (anilpr@iitm.ac.in)}\\  
    
\vspace*{4mm} \noindent
\textbf{ABSTRACT} \\
\bigskip
\begin{minipage}{\textwidth}
\fontsize{10pt}{14pt}\selectfont{

Droplet microfluidics combined with optical detection has emerged as a transformative avenue for high-throughput single-cell assays. However, these platforms often suffer from limited sensitivity and signal heterogeneity due to intrinsic optical and geometrical constraints. We investigate how key operating parameters influence the performance of a droplet-based optomicrofluidic system. Our experiments probe optical interactions between guided light and aqueous droplets encapsulating fluorescent (FL) particles flowing in an oil medium. Geometrical optics simulations model light-droplet interaction, while FL simulations quantify signal variation due to particle size and position. Experimentally, we observe two refracted optical signals: a droplet-refracted signal (DRS) that scales with droplet diameter, and a particle-refracted signal (PRS) arising from light interaction with encapsulated particles. Both experiments and simulations reveal that PRS is prominent when the particle-to-droplet size ratio (\( D^*_\text{p} \)) is between 0.23-0.33, enabling label-free detection. Experiments also show that particles located near the droplet center (\( r^*_\text{p} < 0.4 \)) exhibit a reduced angular dependence, producing more uniform FL signals. Simulations indicate that the FL intensity increases with $ D^*_\text{p}$, with a strong rise from 0.33 to 0.5 and a more gradual increase up to 0.66. Further, simulations show that reducing the thickness of the oil layer consistently enhances fluorescence by minimizing optical losses at the droplet–channel interface. Our results demonstrate that controlling \( D^*_\text{p} \), particle position, and oil layer thickness significantly improve FL strength and uniformity. These findings establish a quantitative and experimentally validated framework for optimizing droplet-based fluorescence detection, with potential applications in microflow cytometry and single-cell assays.

}
\end{minipage}
\vspace{1.5em}
\end{@twocolumnfalse} 
]

{

\setstretch{1.0}

\section{Introduction}\label{sec:1}

Droplet microfluidics has emerged as a powerful platform for high-throughput biochemical screening, enabling the generation of monodisperse droplets ranging from picoliters to microliters at rates up to several MHz, while minimizing sample and reagent consumption~\cite{Hsieh2021, Yu2021, Wang2020, Guo2012, Seemann2011}. Each droplet functions as an independent microreactor, allowing precise control, scalability, and prevention of cross-contamination~\cite{Gupta2024, Hess2019}. Accurate analysis of droplet contents is essential, and optical detection techniques are particularly attractive due to their non-invasive, real-time, and high-throughput capabilities~\cite{Caen2018, Prnamets2021}. While flow cytometry remains the benchmark for single-cell analysis in clinical diagnostics~\cite{Panwar2023}, the integration of droplet microfluidics with optical detection expands its applicability to areas such as single-cell profiling, sequencing, drug discovery, and pathogen detection~\cite{Huang2022}. Nevertheless, conventional optofluidic setups often rely on bulky and expensive instrumentation, requiring precise alignment and technical expertise, which poses barriers to widespread adoption in standard laboratory settings~\cite{Myers2008}.

To overcome these challenges, optomicrofluidic systems integrate optical fibers into microfluidic chips, creating simplified, flexible, and cost-effective setups \cite{Hengoju2022}. These fibers serve as light guides and enable excitation and collection of fluorescence, absorbance, and scattered light signals. The time-shift technique measures temporal delays in scattered light and facilitates precise analysis of particle size, velocity, and refractive index \cite{Schfer2014}. Efforts to enhance optomicrofluidic system performance include various modifications: integrating microlenses \cite{Seo2004, Llobera2007}, collimators, multi-level chip designs, and channel dimension adjustments \cite{Gielen2016}, optical fibers with varied numerical apertures and lensed fibers \cite{Mohan2020}. Similarly, focused forward scattering signals have demonstrated enhanced detection by altering the refractive indices of the droplet and the medium \cite{Liang2024PCR, Liang2024}.

Accurate measurements of optical signals are critical to the success of optomicrofluidic detection, directly impacting the sensitivity, specificity, and reliability of droplet-based biochemical assays \cite{Li2019, Gao2023, Huang2014,Tavakoli2019,Kumari2023}. However, system-induced variations in optical signals can significantly compromise measurement accuracy. In a recent study, five distinct microparticle populations, each labeled with increasing concentrations of fluorophore, exhibited substantial standard deviations in fluorescence intensity within an optofluidic platform \cite{Gupta2024}, highlighting pronounced signal variations even among well-defined populations. Despite its critical impact on assay fidelity, fluorescence intensity variation in particle-encapsulated droplets within fiber-integrated microfluidic systems remains largely unaddressed.

\begin{figure*}[t]
\centering
\begin{minipage}[b]{\textwidth}
    \centering
    \includegraphics[width=\linewidth]{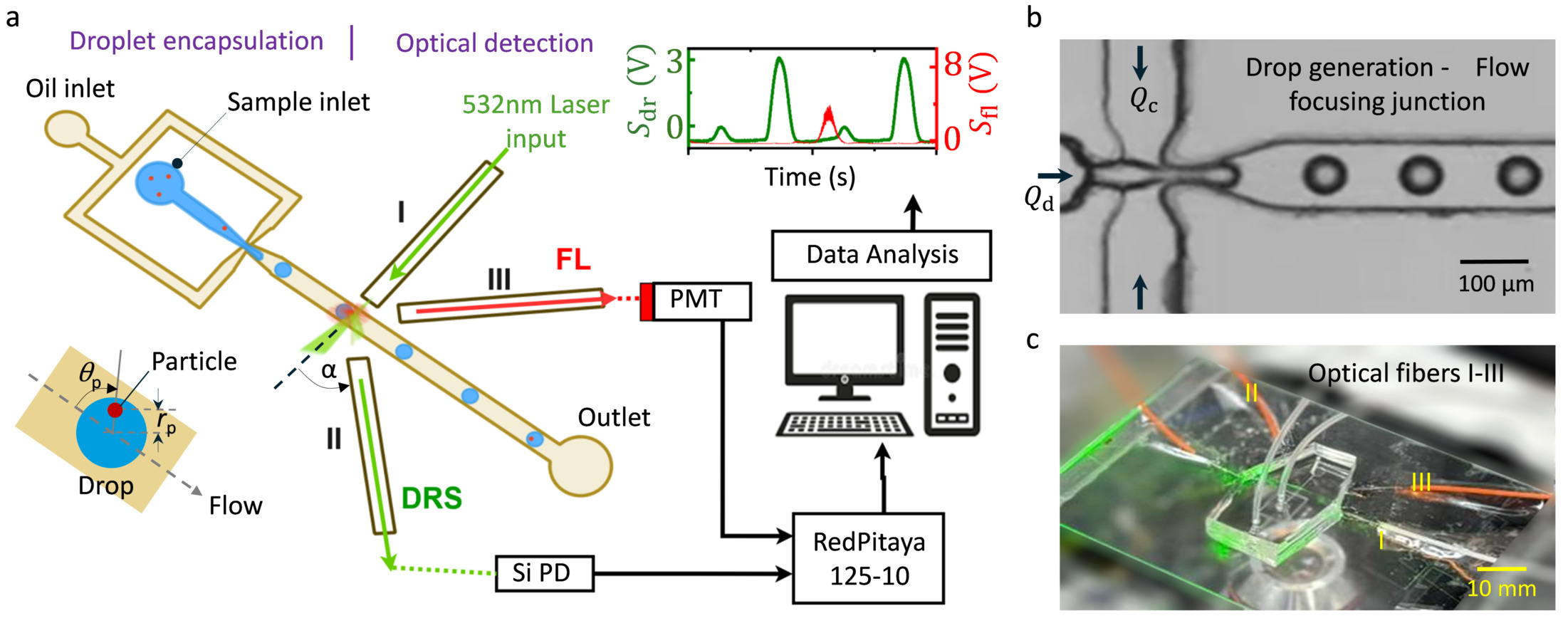}     
\end{minipage}
\caption{(a) Schematic of the optomicrofluidic experimental setup showing the droplet encapsulation and optical detection regions, (b) Experimental image of the flow focusing junction used for droplet generation, (c) Photograph of the optomicrofluidic device with the excitation and collection fibers.}
\label{FIG:1}
\end{figure*}

The interaction of light with drop-encapsulated microparticles involves complex scattering, absorption, and refraction phenomena, which are not yet fully understood. Addressing these optical complexities is vital for improving the accuracy and sensitivity of fluorescence-based detection, particularly in applications requiring high-fidelity single-particle analyses. In particular, the fluorescence collection efficiency is influenced by the droplet curvature\cite{Koegl2020}, which affects light absorption, scattering, and refraction, thereby altering the spatial intensity distribution of the emitted signal. A deeper understanding of system parameters, such as the particle-to-droplet size ratio, particle position within the droplet, and particle angular orientation relative to the optical axis, is essential for improving fluorescence detection in fiber-integrated droplet microfluidic platforms. Such insights will be key to advancing optomicrofluidic measurements targeting quantitative studies on cellular heterogeneity.

In this work, we investigate how operating parameters influence the performance of a droplet based optomicrofluidic system using both experiments and simulations. Experimentally, we generate particle encapsulated droplets and measure optical signals, including droplet refracted and particle reflected or scattered signals, along with fluorescence emission, using fiber integrated detection. Complementing these measurements, simulations based on geometrical optics and fluorescence modeling quantify light droplet interactions and signal variation. We systematically examine the effects of droplet size, particle to droplet size ratio, particle position, including angular orientation and oil layer thickness on fluorescence signal characteristics. This framework supports the design of improved optomicrofluidic systems for microflow cytometry and single cell analysis.

\section{Experimental details}\label{sec:exp}

 \subsection{Device Geometry}\label{device fabrication}
The optomicrofluidic device comprises two main regions: a droplet encapsulation region with a flow-focusing geometry and an optical detection region (Fig.~\ref{FIG:1}a). The microchannel has a depth of 130~ µ m throughout, narrowing to 30~ µ m at the throat for droplet generation and expanding to 100~ µ m in the main channel (Fig.~\ref{FIG:1}b). A photograph of the device is shown in Fig.~\ref{FIG:1}c. The detection region includes three grooves (I–III) to hold standard optical fibers (OD 125~ µ m). Each groove is 130~µ m wide and deep, placed 100~ µ m from the fluid channel wall. Groove I ($\alpha=0^\circ$) holds the excitation fiber. Groove II ($\alpha=45^\circ$) collects side-scattered signals, and Groove III ($\alpha=135^\circ$) collects fluorescence signals. Fabrication involves standard photolithography to create a silicon master, followed by PDMS soft lithography using molding and bonding. Additional fabrication details are available elsewhere~\cite{Sajeesh2014}.

\subsection{Materials}\label{materials}
Fluorescent microparticles (15~ µ m, microParticles GmbH, Germany) are suspended in deionized water (resistivity: 18.2~M$\Omega\cdot$cm, ELGA, UK) with 5\% Pluronic F127 (Sigma-Aldrich, India) to prevent agglomeration. The suspension is washed twice and sonicated for 5~minutes using an ultrasonic cleaner (Maxsell, India). The continuous phase is mineral oil (Sigma-Aldrich, India) with 5\% (wt/wt) Span 85 added for droplet stability and filtered using 0.2~ µ m PTFE filters to remove unwanted particles.

\begin{figure*}[t]
\centering
\begin{minipage}[b]{\textwidth}
    \centering
    \includegraphics[width=0.8\linewidth]{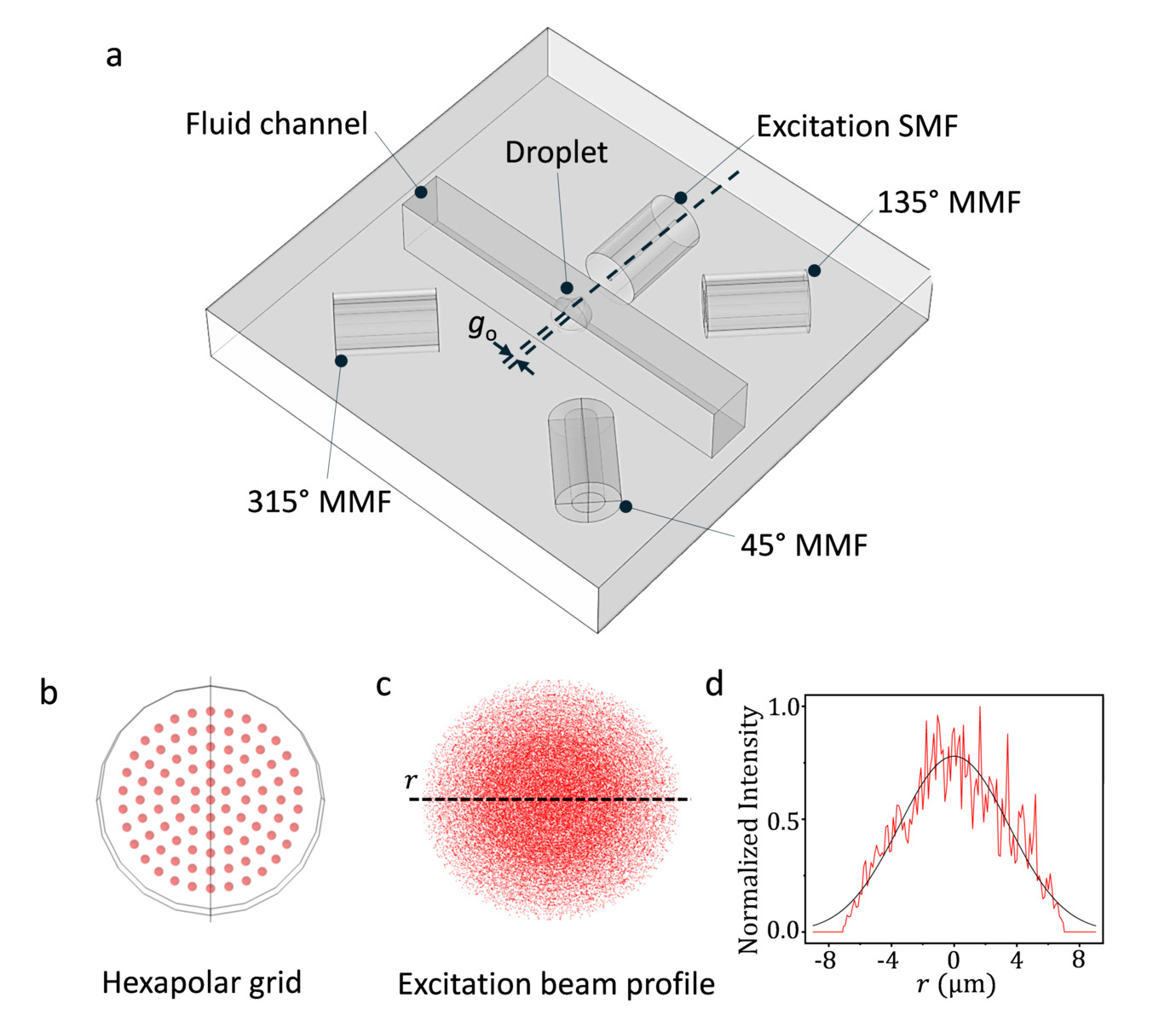}     
\end{minipage}
\caption{(a) Model geometry of the optomicrofluidic device with a droplet, fluid channel, and excitation/collection optical fibers used for the numerical simulations, (b) Model showing the positions of ray release from a hexapolar grid within the core of a singlemode fiber (SMF),  (c) Geometry of the excitation beam profile which is measured to be 7.82  µ m, (d).  Excitation profile plotted across the optical beam, fit to a Gaussian $f(x) = 0.78~e^{-r^2/(2\sigma^2)}$ where $\sigma = 3.49 $ with $R^2 = 0.88$.}
\label{FIG:2}
\end{figure*}

\subsection{Experimental Setup}
The experimental setup is shown in Fig.~\ref{FIG:1}a; see the Supplementary Material \cite{Supplementary_Material} for photograph . Experiments are performed on an inverted microscope (IX73, Olympus Inc., Germany) with a high-speed camera (Phantom v2640, Vision Research Inc., USA). The microparticle suspension and mineral oil are infused using syringe pumps (neMESYS, Cetoni, Germany) via polyethylene tubing (BTPE-50, Instech Laboratories, USA). Single-mode (9~ µ m core) and multimode (62.5~ µ m core) optical fibers are stripped, cleaved flat, and inserted into the grooves filled with index-matching fluid (Fiber Instruments Sales Inc., USA). A 532~nm fiber-coupled laser (Wave Form Systems, USA) operating at 1~mW excites the droplets via the single-mode fiber. Multimode fibers placed at 45$^\circ$ and 135$^\circ$ collect side-scattered and fluorescence signals, respectively. Signals are directed to a Si photodiode (DET02AFC/M, Thorlabs) and a photomultiplier tube (PMT2101/M, Thorlabs), with a 575/50~nm bandpass filter (Chroma Technology, USA) for fluorescence. Data acquisition is handled by STEMlab 125-10 kit (RedPitaya, Europe).

\subsection{Data Processing and Video Analysis}\label{video analysis}
Voltage signals from the photodiode (PD) and Photo multiplier tube (PMT) are acquired by RedPitaya at a fixed sampling rate based on flow conditions and recorded as CSV files using Jupyter Notebook. Time-series data are analyzed in Python to determine peak height ($S_\text{dr}$), width at 90\% ($t_\text{dr}$), inter-peak distance, and frequency ($f_\text{dr}$) for droplet-refracted signals. 
Fluorescence peaks ($S_\text{fl}$) are similarly extracted. Based on fluorescence signal images, incomplete peaks resulting from limitation in data acquisition with STEMlab 125-10, are excluded. Videos are processed in ImageJ to measure droplet diameter and particle position. The centroid of each droplet and encapsulated particle is identified in the frame prior to light interaction. The distance ($r_p$) and angle ($\theta_p$) between the droplet and particle centroids are computed as shown in

\section{Numerical Model}

\subsection{Light–Droplet Interaction}\label{light_droplet_interaction}

Light interaction with droplets plays a vital role in optomicrofluidic detection systems~\cite{Zhu2019}. When the droplet diameter ($D_\text{d}$) is much larger than the excitation wavelength ($\lambda$), i.e., $D_\text{d}\gg \lambda$, geometric optics is applicable, and scattering is predominantly governed by geometric effects~\cite{Houghton1949}. The size parameter $\alpha = \pi D_\text{d}/\lambda$ characterizes this regime, with $\alpha \gg 1$ indicating dominance of geometric scattering. Under these conditions, the droplet behaves like a spherical object that reflects and refracts light in multiple directions, while diffraction effects remain minimal~\cite{Li2019_sim}. Light propagation is governed by Snell’s law~\cite{Born1999}.

Simulations are conducted using the Ray Optics module in COMSOL Multiphysics 6.1~\cite{Vinothkumar2016}. For inhomogeneous media, ray propagation is described by:
\begin{equation}
\frac{dq}{dt} = \frac{k}{n}, \quad \frac{dk}{dt} = -\nabla n(q),
\label{eq1}
\end{equation}
where $q$ is the position vector and $k$ the wave vector, and $n$ denotes the refractive index of the medium~\cite{comsolRayOptics2020}. Although Eq.~\ref{eq1} is written in a general form applicable to inhomogeneous media, in the present simulations each material domain is treated as optically homogeneous, with a constant refractive index assigned within that domain. During ray propagation, the refractive index is updated when a ray crosses an interface between different materials, and refraction is computed using Snell’s law to ensure phase continuity. 

The simulation geometry of the optomicrofluidic device, described in Section~\ref{device fabrication}, is illustrated in Fig.~\ref{FIG:2}(a). Rays released from hexapolar grid within the core of a single-mode fiber (SMF), as shown in Fig.~\ref{FIG:2}(b).  The excitation beam’s spot radius at the channel center, in the absence of droplets, is estimated to be 7.82  µ m (Fig.~\ref{FIG:2}(c)) using ray optics simulation and corresponding intensity profile as a function of radius is shown in Fig.~\ref{FIG:2}(d).
The simulation computes ray intensities using Stokes parameters to account for polarization~\cite{Schaefer2007}. "Freeze wall" boundaries accumulate incident intensities normalized by surface area. Secondary rays from internal reflections are managed using intensity thresholds and release limits for computational efficiency. Output rays collected at multimode fibers placed at 45° and 315° represent droplet-refracted signals. Parametric sweep over the droplet position (denoted by offset $g_o$, as marked in Fig.~\ref{FIG:2}a) enables dynamic modeling of light-droplet interaction during droplet transit through the detection region.

 \begin{figure*}[t]
\centering
\begin{minipage}[b]{\textwidth}
    \centering
    \includegraphics[width=\linewidth]{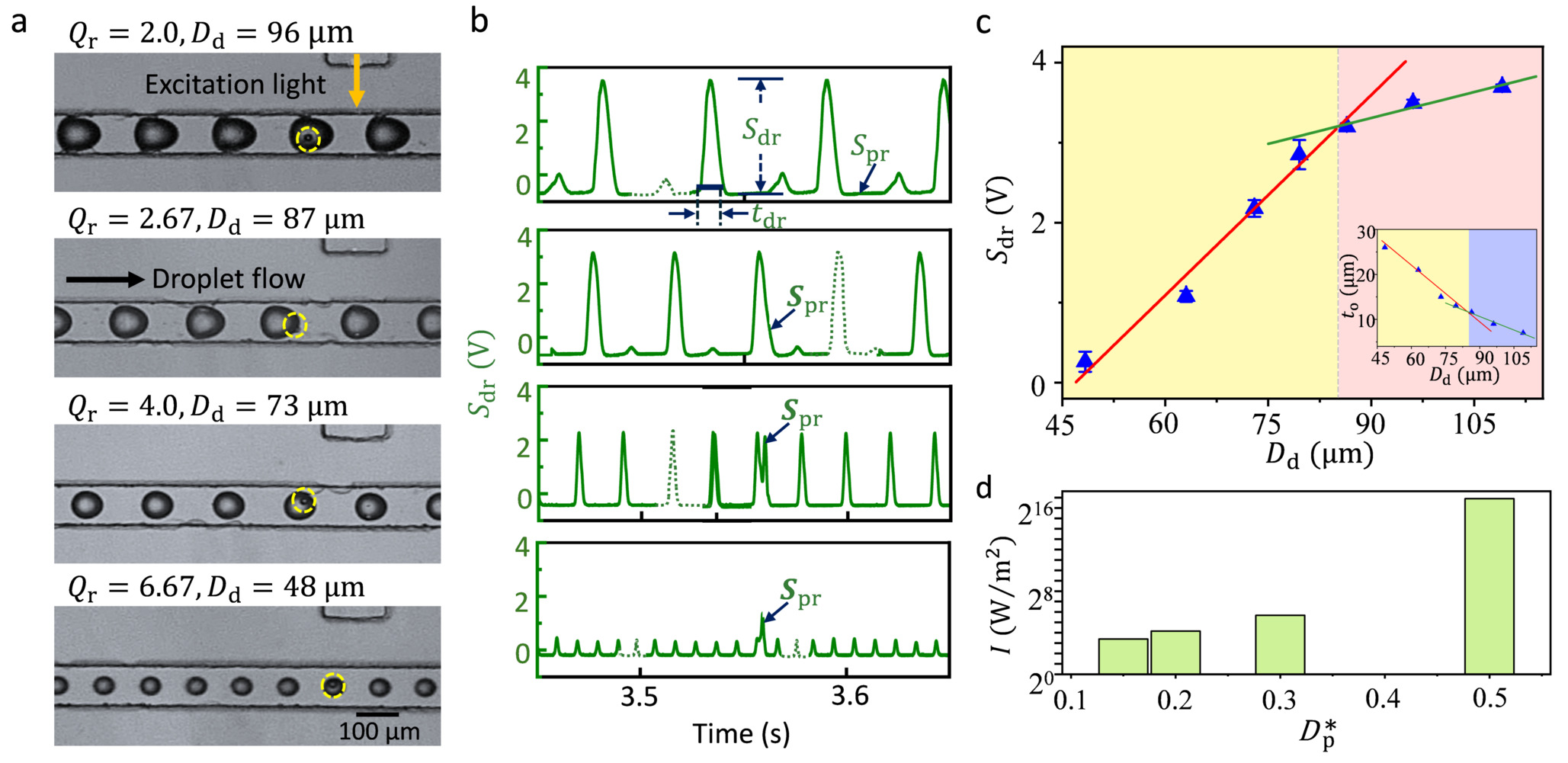}
        
\end{minipage}
\caption{(a) Images of the droplets generated at varying $Q_\text{r}$, (b) corresponding droplet refracted signal acquired. The dashed line indicates missing data due to real-time data acquisition limitations, (c) Correlation plot for droplet refracted signal height with droplet diameter. Inset: Oil layer thickness decreases with increasing droplet diameter. Error bars indicate ±SD. (d) Simulation studies show the particle-refracted signal increases with increasing \( D^*_{\text{p}} \), i.e., smaller droplet size at $\theta_{\text{p}} = 90\degree$ with fixed particle diameter of {15} { µ m} and a fixed 1.5  µ m gap between the particle surface and the droplet boundary.}
\label{FIG:3}
\end{figure*}

\subsection{Fluorescence Simulation}\label{fluorescence simulation}

Fluorescence occurs when fluorophores absorb excitation light and emit light at longer wavelengths~\cite{Haustein2007}. In droplet-based systems, multiple light interactions—absorption, scattering, and internal reflection—affect fluorescence behavior~\cite{Chan2012,Cotterell2022}. Enhancing signal strength and uniformity is critical for high-throughput screening and single-cell studies~\cite{Liang2024PCR, Persichetti2017}.

A two-step computational approach is implemented using COMSOL~\cite{Soorya2014, singh2011}. First, ray tracing simulations calculate the excitation light reaching the particle surface. Second, in COMSOL fluorescence emission is modeled using the Helmholtz diffusion equation:

\begin{equation}
\nabla(-c\nabla u)+au = f 
\end{equation}
with parameters,
\begin{equation*}
    u = \phi 
\end{equation*} 
\begin{equation*}
    c = D= \frac{1}{3(\mu_a+\mu_s^{'})} 
\end{equation*} 
\begin{equation*}
    a = \mu_a
\end{equation*} 
\begin{equation*}
    f = S
\end{equation*} 

where $u$ is the fluorescence intensity, $D$ is the optical diffusion coefficient, $\mu_{a}$ and $\mu_{s}^{'}$ are absorption coefficient and reduced scattering coefficient at emission wavelengths, respectively,  and $S$ is the local excitation intensity which is computed from ray optics simulation.  $\mu_{a}$ and $\mu_{s}^{'}$ for PDMS, mineral oil and water were taken from the literature \cite{Greening2014, Lapsley2011, MarnSerrano2019, Przybylek2023, AguilarArevalo2009, Gokhale2021, Buiteveld1994, Pope1997}; The corresponding values are provided in the Supplementary Material \cite{Supplementary_Material}.

Fluorescence output is computed by integrating the emission over the MMF core area at 135°, providing the simulated fluorescence intensity. Particle scattering is also modeled using a point source at the particle center emitting isotropically. The scattered light collected at MMFs is weighted by the surface excitation profile to determine the particle’s refracted signal. This integrated model enables realistic simulation of fluorescence and refracted signals in droplet-based optomicrofluidic systems.

\section{Results and Discussion}

\subsection{Effect of droplet size on droplet and particle refracted signals}\label{DRS}

Water-in-oil droplets are formed by infusing DI water, containing 0.1\% Tween 20, 15\% Optiprep, and ~~~{$3 \times 10^5$} ml$^{-1}$ of 15  µ m pluronic-treated microparticles, as the dispersed phase, and mineral oil with 5\% Span 85 as the continuous phase. Droplet size is controlled by varying the continuous phase flow rate ($Q_\text{c}$) while maintaining a constant $Q_\text{d} = 0.3$ µ ml/min, dispersed phase flow rate. Images of droplets at different flow rate ratios ($Q_\text{r} = Q_\text{c}/Q_\text{d}$) are shown in Fig.~\ref{FIG:3}(a). As $Q_\text{r}$ increases, droplet diameter ($D_\text{d}$) decreases, with a minimum stable size of 48  µ m (see the Supplementary Material \cite{Supplementary_Material}).

\begin{figure*}[t]
\centering
\begin{minipage}[b]{\textwidth}
    \centering
    \includegraphics[width=\linewidth]{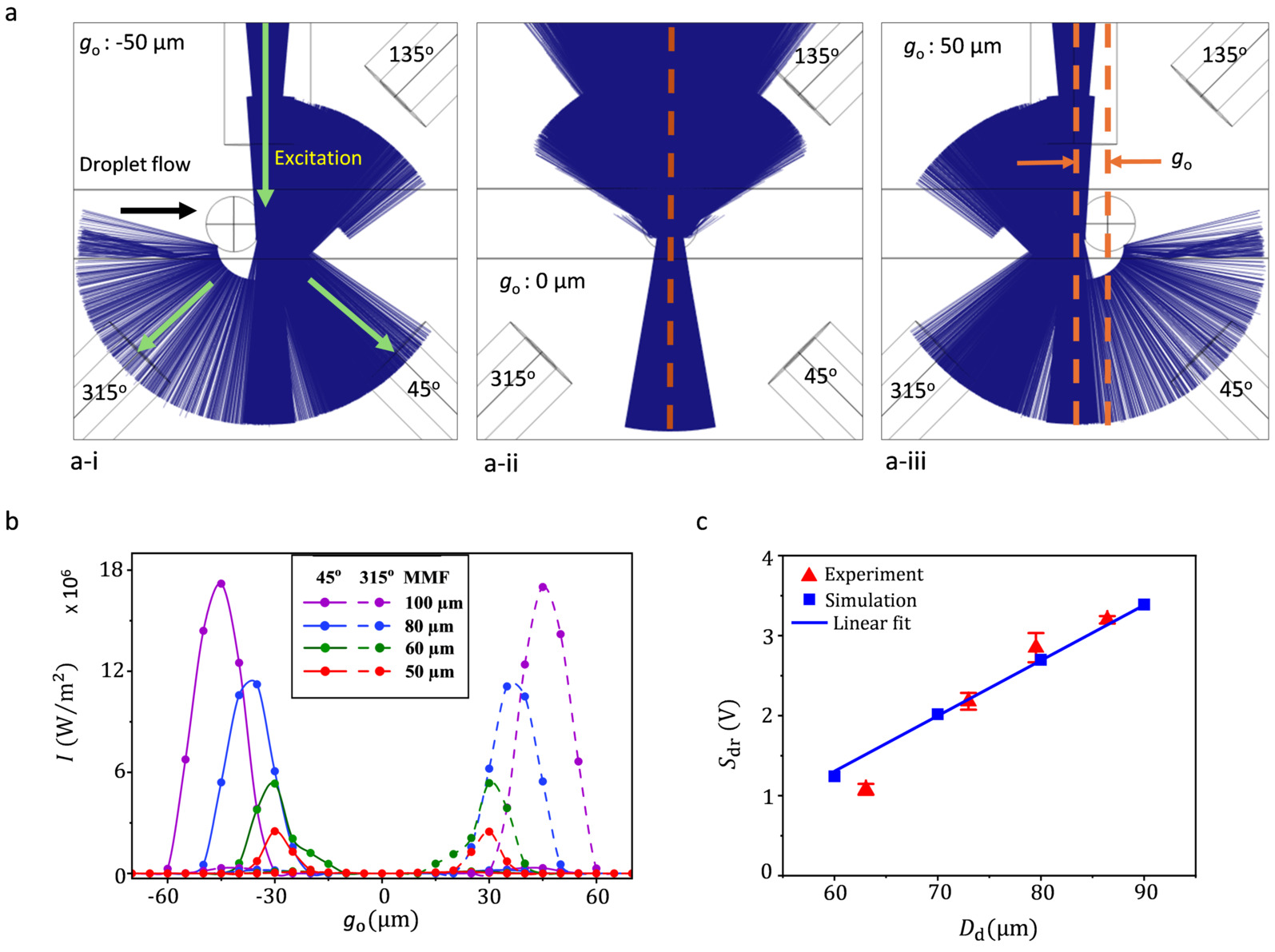}
        
\end{minipage}
\caption{Comparison of experimental and simulated droplet-refracted signals at varied positions. (a-i) Ray trajectory of the excitation source at droplet position $g_\text{o}$ = -50  µ m, (a-ii) Ray trajectory at $g_\text{o}$  = 0  µ m, (a-iii) Ray trajectory at $g_\text{o}$  = 50  µ m, (b) Average accumulated intensity collected at 45$\degree$ (solid line) and 315$\degree$ MMF (dash line) for different droplet sizes, (c) Comparison of droplet refracted signals for varying droplet sizes between simulation and experiments. Experimental data points and simulated values exhibit linear and parallel trends indicating consistent scaling behavior. A calibration constant $k_\text{PD} = 2.4 \times 10^7$ was applied using the relation ${S_{\text{dr}}} = k_\text{PD} {I_{{\text{sim}}}}$. The blue line represents a linear fit to the simulated data using the equation $y = Ax + B$, where $A = 0.087$ and $B = -4.2$.  Error bars indicate ±SD. }
\label{FIG:4}
\end{figure*}

When a particle-laden droplet enters the optical region, the laser beam is reflected and refracted, producing signals that encode droplet size and generation rate. Light is collected by a multimode fiber (MMF) at $\alpha = 45^\circ$ (Fig.~\ref{FIG:1}a, ~\ref{FIG:1}c, and Supplementary Material \cite{Supplementary_Material}). A high-speed camera captures droplet formation, while a STEMlab 125-10 system records the droplet-refracted signal (DRS). Representative DRS waveforms for different $Q_\text{r}$ values are shown in Fig.~\ref{FIG:3}(b). Key DRS features, peak height ($S_\text{dr}$), peak width ($t_\text{dr}$) at 90\% of peak height, and droplet generation rate ($f_\text{dr}$), are extracted. Peak height ($S_\text{dr}$), influenced by droplet size, is the most reliable sizing metric, as $t_\text{dr}$ depends on excitation beam width and continuous phase velocity ($U_\text{c}$), limiting its utility \cite{Schfer2014}. Plots of $S_\text{dr}$ and $t_\text{dr}$ versus $D_\text{d}$ are shown in Fig.~\ref{FIG:3}(c) and see the Supplementary Material \cite{Supplementary_Material}. Both metrics decrease as $D_\text{d}$ reduces with increasing $Q_\text{r}$. In Fig.~\ref{FIG:3}(c), $S_\text{dr}$ exhibits two regimes: in the large-droplet regime ($D_\text{d} > 85$  µ m), it decreases gradually with droplet size, whereas in the small-droplet regime ($D_\text{d} < 85$  µ m), it decreases at a higher rate. Analysis of Covariance (ANCOVA) reveals a statistically significant (p = 0.024) difference in the droplet-size-dependent slopes of peak height between the two regimes. Larger droplets deform against channel walls, while smaller droplets retain a near-spherical shape. This shift in morphology affects optical behavior. In large droplets, oil layer thickness, \(t_\text{o}\) = (\(W_\text{c}\)-\(D_\text{d}\))/2, where \(W_\text{c}\) is the channel width and $D_\text{d}$ is the diameter of droplet, increases gradually. In small droplets, \(t_\text{o}\) increases sharply (Fig.~\ref{FIG:3}(c) inset), altering light paths and enhancing scattering into the oil. As a result, signal intensity falls, highlighting sensitivity to droplet shape and confinement.

A secondary peak, termed the particle-refracted signal (PRS), detected alongside the droplet-refracted signal (DRS) as it approaches the MMF at $\alpha = 45^\circ$. The PRS originates from light refracted and scattered by encapsulated particles, and its peak amplitude ($S_\text{pr}$) becomes more prominent with smaller droplets (Fig.~\ref{FIG:3}b). To understand this behavior, simulations were conducted to analyze the influence of droplet size on PRS intensity while keeping the particle size fixed. These simulations focused on the optical response at the particle location, allowing isolation of scattering effects due to spatial confinement. Fig.~\ref{FIG:3}(d) shows the simulation results for a constant particle size, $D_\text{p} = 15$  µ m and decreasing droplet size. The refracted signal intensity collected at the particle position increases with the particle-to-droplet size ratio, $D^*_\text{p} = D_\text{p}/D_\text{d}$. This trend highlights enhanced light confinement and scattering efficiency as the particle occupies a larger relative volume within smaller droplets. These findings support the observed increase in PRS with decreasing droplet size in experimental studies. One question arises: how do these optical responses, arising from light-droplet interactions, change with droplet position along the channel relative to the optical beam path? To further investigate light–droplet interactions, we perform numerical simulations in the following section.

\subsection{Modeling the impact of droplet size on refracted signal}

Numerical simulations are performed using the Ray Optics module in COMSOL Multiphysics 6.1 \cite{Vinothkumar2016}, with material refractive indices defined accordingly. The model, as outlined in Section \ref{light_droplet_interaction}, evaluates the optical signal characteristics as a function of droplet size and axial position. To improve signal capture, a second multimode fiber (MMF) is introduced at $\alpha = 315^\circ$, complementing the primary MMF at $\alpha = 45^\circ$. The excitation beam’s spot radius at the channel center, in the absence of droplets, is estimated to be 7.82  µ m (Fig.~\ref{FIG:2}(c)) using ray optics simulation. To assess signal variation due to droplet position, the axial offset between the droplet center and beam center ($g_\text{o}$) is varied in 5  µ m increments. 

Fig.~\ref{FIG:4}(a) shows ray trajectories at different $g_\text{o}$ values for a droplet diameter of 80  µ m. At $g_\text{o} = -50$  µ m (Fig.~\ref{FIG:4}(a)-i), the droplet begins to intersect the beam path, resulting in strong refraction toward the $\alpha = 45^\circ$ MMF. At $g_\text{o} = 0$  µ m (Fig.~\ref{FIG:4}(a)-ii), the beam passes perpendicularly through the droplet, creating two distinct reflection zones at the fluid–PDMS interfaces. As the droplet exits the optical region at $g_\text{o} = 50$  µ m (Fig.~\ref{FIG:4}(a)-iii), refraction is directed toward the $\alpha = 315^\circ$ MMF. This leads to two signal peaks: a leading peak at $45^\circ$ as the droplet enters and a lagging peak at $315^\circ$ as it exits \cite{Gupta2024}. The separation distance and intensity of these peaks depend on droplet size, refractive index, and velocity \cite{Schfer2014}. Here, the average accumulated intensity ($I$), defined as energy per unit time per unit area at the MMF core, is used to quantify this response. Fig.~\ref{FIG:4}(b) shows $I$ profiles at both detectors ($\alpha$\ = \(45^\circ\) and \(315^\circ\)) for droplets of varying sizes. Peak intensity increases with droplet size, while the separation between leading and lagging peaks also widens. Since beam width and droplet velocity were held constant in simulations, peak width remained nearly constant (40  µ m), making peak height a more reliable indicator of droplet size than width.

The numerical model validation is shown in Fig.~\ref{FIG:4}(c), comparing simulated peak heights with experimental measurements. Both show a linear correlation between peak height and droplet size, with minimal deviations \cite{Wu2016}. These results confirm the accuracy of the ray-optics model in capturing size-dependent optical behavior. While DRS intensity increases with increasing droplet radius, fluorescence collection efficiency is also influenced by encapsulation dynamics. How do droplet encapsulation characteristics influence the excitation and collection of fluorescence? These aspects are explored in detail in the following sections.

\subsection{Effect of droplet encapsulation on fluorescence signal}

Optomicrofluidic detection of droplet-encapsulated microparticles or cells primarily aims to measure variations in particle properties \cite{Lu2016}, such as heterogeneous surface marker expression \cite{Behsen2024}, by analyzing changes in fluorescence signal. However, variations in fluorescence can also arise from differences in droplet encapsulation parameters, complicating interpretation and limiting the accuracy of applications that require precise quantification or differentiation between similar cell subpopulations. Understanding and minimizing fluorescence signal heterogeneity due to encapsulation effects is therefore critical. To investigate the impact of encapsulation parameters on fluorescence, we analyze experimental fluorescence intensity data from ~~~~~15  µ m fluorescent particles encapsulated in droplets of varying sizes.

\begin{figure*}[!t]
\centering
\begin{minipage}[b]{\textwidth}
    \centering
    \includegraphics[width=\linewidth]{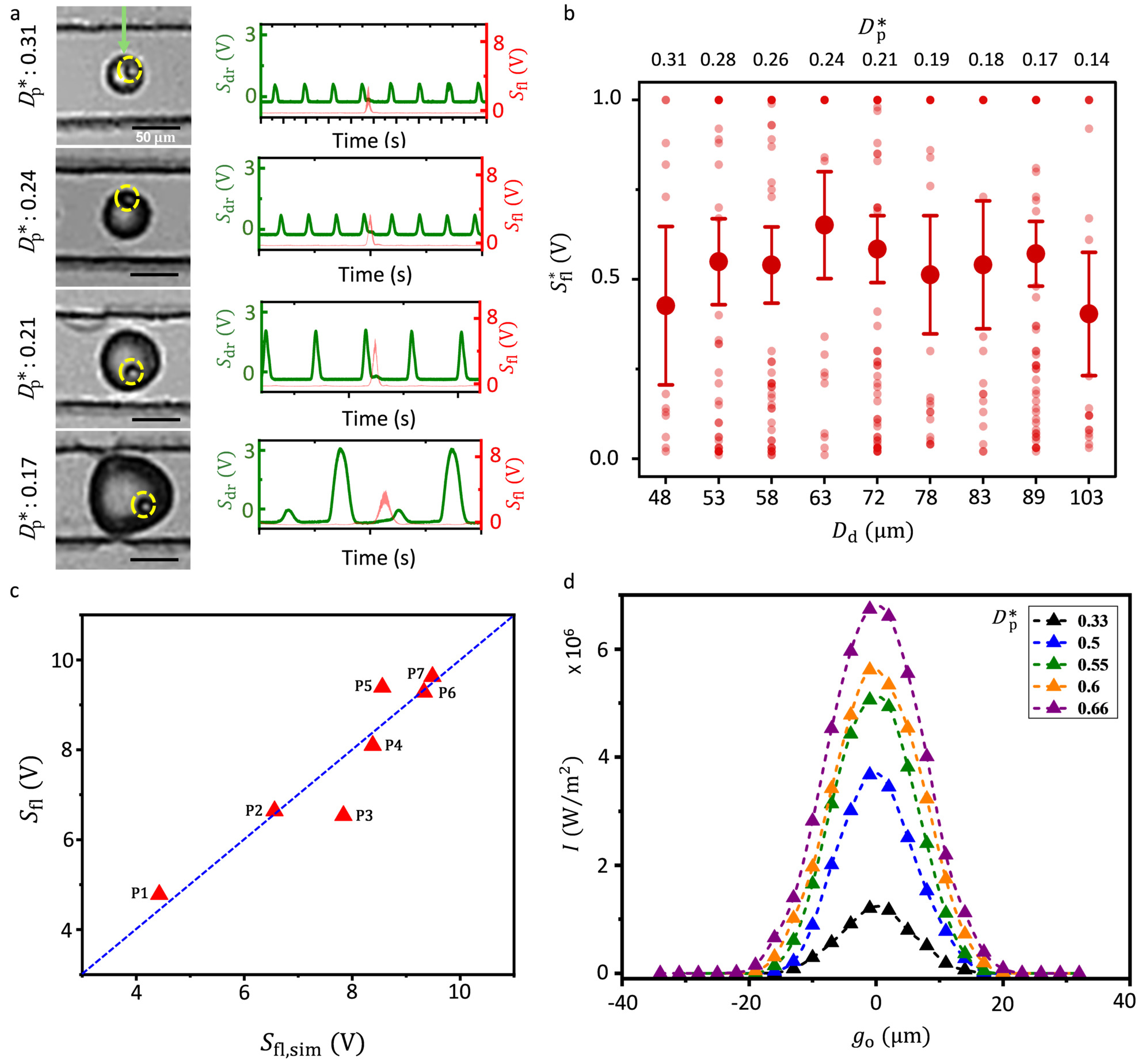}  

\end{minipage}
\caption{Effect of droplet size and particle confinement on fluorescence.(a) Particle encapsulated droplets of varying size and corresponding droplet refracted signal (green) and particle fluorescence signal (red), (b) Normalized fluorescence intensity values for varying particle-to-droplet size ratio, (c) Comparison of simulated and experimental particle fluorescence for different particle configuration (P1-P7).  Each data point corresponds to a unique droplet–particle configuration; see the Supplementary Material \cite{Supplementary_Material} for $D_{\text{d}}$, $r_{\text{p}}$ and $\theta_{\text{p}}$ values. A calibration constant ${k_{\text{PMT}} = 3.01 \times 10^{-6}}$ was applied using the relation ${S_{\text{fl},{\text{sim}}}} = k_{\text{PMT}} {I_{{\text{fl}},{\text{sim}}}}$, (d) Simulated fluorescence intensity from 135$\degree$ MMF for varied particle-to-droplet size ratio with particle positioned at $\theta_{\text{p}} = 90\degree$ with fixed $D_{\text{d}}$ of {30} { µ m}. 
}
\label{FIG:5}
\end{figure*}

Experimental images of encapsulated fluorescent particles across a range of droplet sizes (i.e., varying $D^*_\text{p}$) and their corresponding droplet refracted and fluorescence signals are shown in Fig.~\ref{FIG:5}(a). From 367 DRS files, droplet diameters were obtained using the correlation plot in Fig.~\ref{FIG:3}(c). Fig.~\ref{FIG:5}(b) shows the variation of normalized fluorescence intensity ($S^*_\text{fl}$) with droplet diameter, where $S_\text{fl}$ is scaled by the STEMlab 125-10 maximum measurable voltage (20.4 V). A wide spread in fluorescence intensity is observed across the $D_\text{d}$ and corresponding $D^*_\text{p}$ range, with no clear trend. This suggests that fluorescence output is not solely dependent on droplet size. Instead, signal heterogeneity likely results from the combined effects of multiple factors such as microparticle size, particle position within the droplet, and oil layer thickness. A detailed understanding of how each parameter contributes is essential to optimize droplet-based optical detection. The following section investigates these individual factors and their influence on fluorescence response.

\subsection{Effect of particle-to-droplet size ratio on fluorescence signal}

The impact of particle-to-droplet size ratio ($D^*_\text{p}$) on fluorescence intensity is examined using ray tracing and fluorescence simulations, keeping droplet diameter and particle position constant. Simulations are based on the model described in Section~\ref{fluorescence simulation}. Model validation, shown in the Fig.~\ref{FIG:5}(c), compares simulated and experimental fluorescence across different droplet configurations. See the Supplementary Material \cite{Supplementary_Material} for details on particle configuration. Simulations were performed using a fixed droplet diameter of 30  µ m, varying particle diameter from 10  µ m to 20  µ m. The particle is located at $\theta_{\text{p}} = 90^\circ$, near the optical axis (see Fig.~\ref{FIG:5}(a) for $D^*_\text{p} = 0.24$). As shown in Fig.~\ref{FIG:5}(d), increasing the $D^*_\text{p}$ from 0.33 to 0.66 results in a pronounced enhancement of fluorescence intensity, corresponding to an approximately fivefold increase. This increase is attributed to the larger particle surface area interacting with a greater fraction of the excitation field, leading to enhanced fluorescence generation and collection efficiency.

The results show that fluorescence intensity increases with $D^*_\text{p}$, although the rate of increase is not uniform. Between $D^*_\text{p}$ = 0.33 and 0.5, the intensity rises sharply, indicating strong size sensitivity in this range, followed by a more gradual increase from 0.5 to 0.66. This behavior suggests diminishing incremental gains in fluorescence for larger particles. Consequently, selecting an appropriate range is important for balancing fluorescence signal strength and size sensitivity. The lower range ($D^*_\text{p}$ =0.33–0.5) offers high sensitivity to particle size, whereas larger $D^*_\text{p}$ values provide stronger absolute signals with reduced sensitivity to small size variations.

\subsection{Effect of microparticle position on fluorescence signal}

\begin{figure*}[!t]
\centering
\begin{minipage}[b]{\textwidth}
    \centering
    \includegraphics[width=0.95\linewidth]{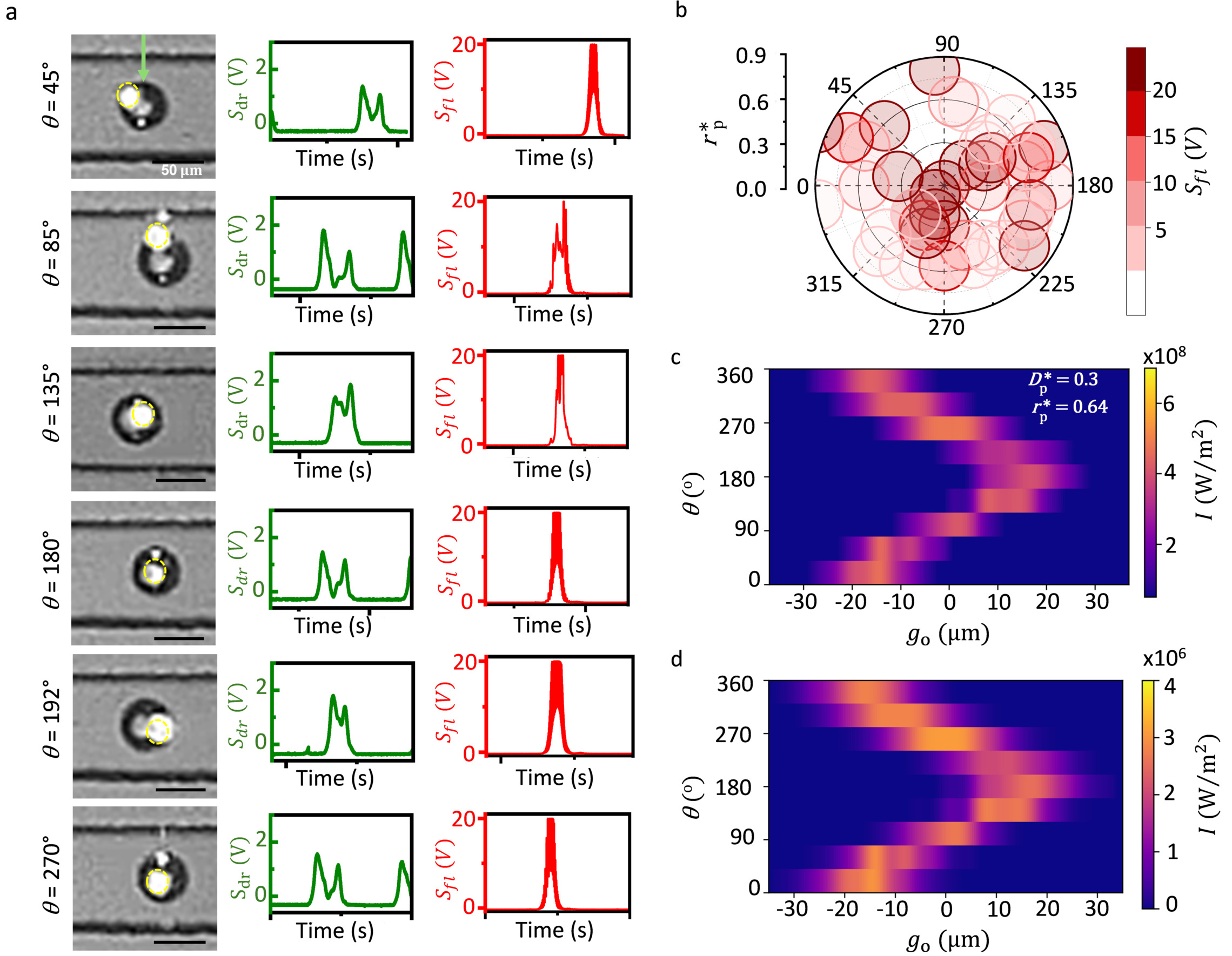}    
\end{minipage}
\caption{ Influence of particle position on fluorescence signal. (a) Particle encapsulated at different position and corresponding droplet refracted signal (green) and particle fluorescence signal (red), (b) Particle position and fluorescence intensity from experimental data for  $D^*_\text{p}$ between 0.27 to 0.3, (c) Simulation for light intensity reaching particle surface positioned at varied angle ($\theta_{\text{p}}$) for fixed $r_\text{p}$ (16  µ m) and $D_\text{d}$ (50  µ m). (d) Simulated fluorescence intensity reaching 135$\degree$ MMF for particle positioned at varied angle ($\theta_{\text{p}}$): 0$\degree$ to 360$\degree$ in steps of 45$\degree$. 
 }
\label{FIG:6}
\end{figure*}

To study the effect of microparticle position on fluorescence signal, we analyze experimental data at a fixed particle-to-droplet size ratio and oil layer thickness. The particle position is described by its radial distance from the droplet center ($r_{\text{p}}$) and angular location ($\theta_{\text{p}}$), measured clockwise from a reference axis (Fig.~\ref{FIG:1}(a)). A total of 237 videos were processed to extract $r_{\text{p}}$ and $\theta_{\text{p}}$, following the method in section~\ref{video analysis}. Data shown in Figs.~\ref{FIG:6}(a) and \ref{FIG:6}(b) correspond to $D^*_{\text{p}} \approx 0.3$ ($\pm$0.03) and oil layer thickness of ~25  µ m ($\pm$2.5  µ m). Fig.~\ref{FIG:6}(a) displays representative particle positions across different angular orientations, together with the corresponding droplet-refracted signal (DRS) and fluorescence profiles, and is intended for qualitative illustration of signal features. The particle location notably affects both the timing and magnitude of the particle-refracted signal (PRS) relative to DRS. Fig.~\ref{FIG:6}(b) presents the corresponding quantitative analysis, showing fluorescence intensity as a function of dimensionless radial position  $r^*_{\text{p}}  = r_{\text{p}}/ R_{\text{d}}$, here $R_{\text{d}}$ is droplet radius. Particle illustrations are shown at half-size for clarity. 
Fluorescence intensity differs significantly between particles near the droplet center and those closer to the boundary, with a nonparametric Mann–Whitney U test confirming higher fluorescence for $r^*_{\text{p}} < 0.4$ than for $r^*_{\text{p}} > 0.4$ ($p < 0.05$).  The fluorescence distributions for the two $r^*_{\text{p}}$ groups are further illustrated using cumulative distribution function (CDF) plot; see the Supplementary Material \cite{Supplementary_Material}. Particles near the center exhibit weak angular dependence, quantified by a coefficient of variation (CV) of approximately 60\%, whereas near-boundary particles show stronger angular dependence (CV $\approx$ 78\%), with enhanced fluorescence near $\theta_{\text{p}} = 135^\circ$ and $270^\circ$.

Across multiple droplets ( \( D^*_{\text{p}} \) ), particles near \(\theta_{\text{p}} \) = \( 135^\circ \) generally exhibit high fluorescence. However, a consistent trend across droplet sizes is absent, likely due to the lack of depth (z-axis) information in position determination, introducing variability. The enhanced signal near \(\theta_{\text{p}} \) = 135$\degree$ is attributed to more efficient collection by the detection fiber placed at \(\alpha \) = \( 135^\circ \).

To further probe angular effects, ray optics simulations were conducted at fixed radial distance \( r_{\text{p}}\) = 16  µ m or \( r^*_{\text{p}} = 0.64 \) for \( D^*_{\text{p}} = 0.3 \). The particle was placed in the beam axis plane at constant depth, and  \(\theta_{\text{p}} \) varied in 45° steps. Incident excitation intensity on the particle surface was computed by freezing rays at the interface. Fig.~\ref{FIG:6}(c) shows light intensity reaching particle, peaks for (\(\theta_{\text{p}} \)) between 270$\degree$ and 315$\degree$, while the detected fluorescence (Fig.~\ref{FIG:6}(d)) peaks near \(\theta_{\text{p}} \) = 135$\degree$, where the collection fiber is positioned. This discrepancy highlights the critical role of spatial alignment between excitation and detection paths. Simulations for varying droplet diameters are provided in the Supplementary Material \cite{Supplementary_Material}.

In summary, for \( D^*_{\text{p}} \) $\approx$ 0.3 ($\pm$0.03) , stronger fluorescence is achieved when particles are either at \( r^*_{\text{p}} < 0.4 \) (any angle) or at \( r^*_{\text{p}} > 0.4 \) but aligned near the collection fiber at \(\theta_{\text{p}} \) = 135$\degree$. Fluorescence variability due to angular position is minimal at lower \( r^*_{\text{p}} \), but increases significantly for particles farther from the center. Thus, maintaining  \( r^*_{\text{p}} < 0.4 \) ensures more consistent fluorescence signals.

\subsection{Effect of oil layer thickness on fluorescence signal}

\begin{figure*}[!t]
\centering
\begin{minipage}[b]{\textwidth}
    \centering  
    \includegraphics[width=0.95\linewidth]{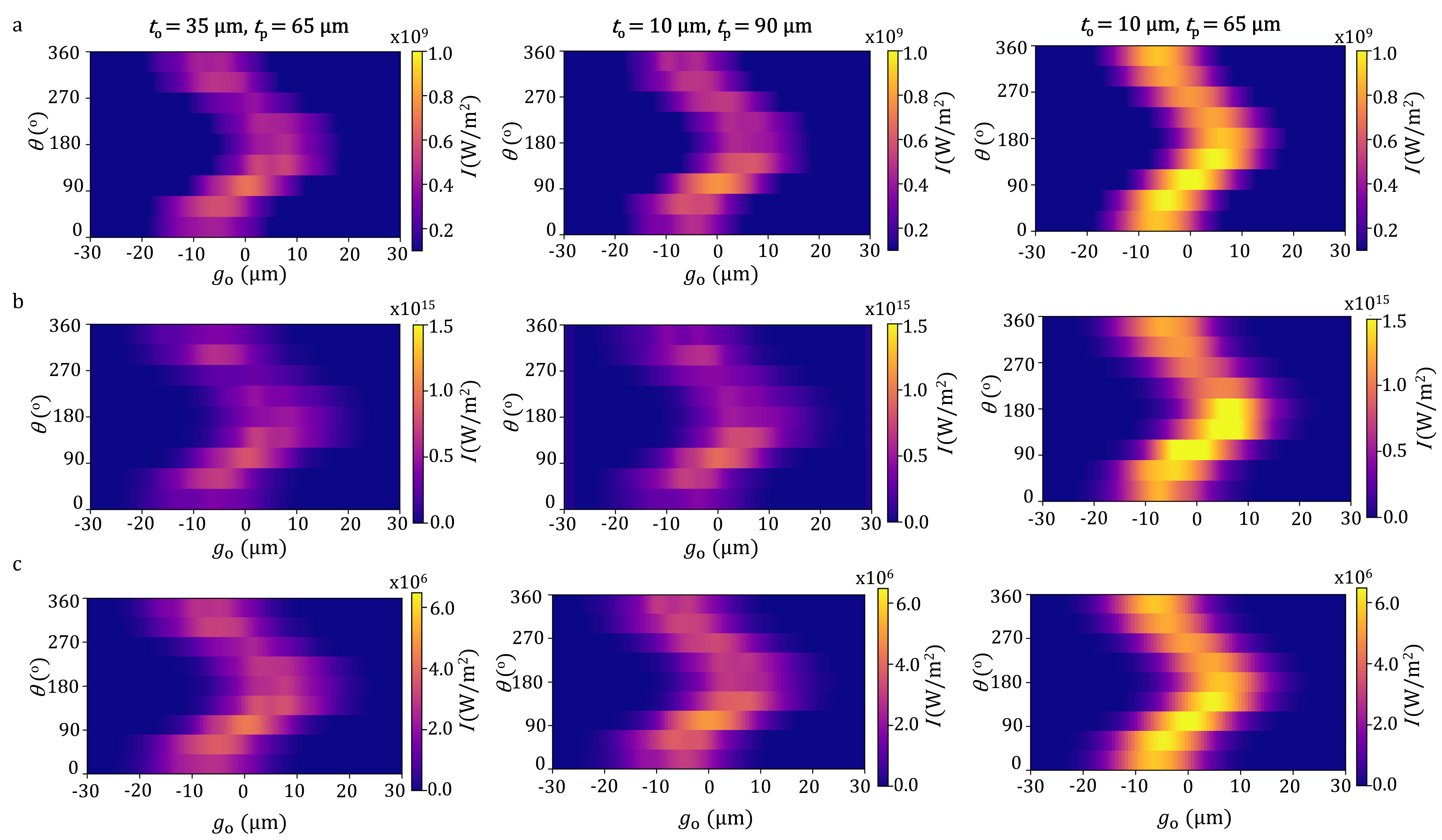}    
\end{minipage}
\caption{
Simulation of oil-layer thickness effects on fluorescence signals for different geometric configurations.
The oil-layer thickness ($t_{\text{o}}$) was varied by adjusting the channel width while keeping the droplet diameter fixed. Rows show (a) excitation intensity at the particle surface, (b) particle scattering intensity reaching the 135$\degree$ multimode fiber (MMF), and (c) fluorescence intensity reaching the 135$\degree$ MMF. The three columns correspond to the baseline geometry, reduced oil-layer thickness with a fixed fiber-to-droplet distance, and reduced oil-layer thickness with a shortened fiber-to-droplet distance, respectively.
}

\label{FIG:7}
\end{figure*}

We used simulations to study the impact of oil layer thickness (\( t_{\text{o}} \)) between the droplet and channel walls on the fluorescence signal, while keeping \( D^*_{\text{p}} \) = 0.5 and particle position fixed (\( r^*_{\text{p}} \) = 0.4). The oil layer thickness was varied by adjusting the channel width \( W_{\text{c}} \), with the droplet diameter held constant at 30~ µ m. Figure~\ref{FIG:7} presents results from two-stage fluorescence simulations, showing the excitation intensity at the particle surface (a), particle scattering intensity (b), and fluorescence intensity (c), arranged in three columns corresponding to different geometric configurations. The left column  (\( t_{\text{o}} \) = 35~ µ m, \( t_{\text{p}} \) = 65~ µ m(thickness of PDMS between SM fiber and the fluid channel wall)) represents the baseline configuration with the largest oil layer. The middle column (\( t_{\text{o}} \) = 10~ µ m, \( t_{\text{p}} \) = 90~ µ m) corresponds to a reduced oil layer thickness while keeping the PDMS thickness constant, thereby maintaining a fixed fiber-to-droplet distance. In contrast, the right column (\( t_{\text{o}} \) = 10~ µ m, \( t_{\text{p}} \) = 65~ µ m) represents a reduced oil layer thickness accompanied by a shorter fiber-to-droplet distance.

Comparing the left and middle columns isolates the effect of oil-layer absorption, showing that reducing  (\( t_{\text{o}} \) leads to an increase in fluorescence intensity of approximately 20\%, despite minimal changes in excitation intensity. This indicates that reduced absorption losses in the oil layer directly contribute to fluorescence enhancement. In contrast, when the reduction in oil-layer thickness is accompanied by a shorter fiber-to-droplet distance, the fluorescence signal exhibits a substantially larger enhancement of approximately 50\%,reflecting the combined contributions of reduced absorption losses and enhanced excitation efficiency. Consistent with this interpretation, the excitation intensity at the particle surface increases markedly in the right column across all angular positions.

Overall, these results demonstrate that minimizing the oil layer thickness improves fluorescence detection through two mechanisms: reduced absorption losses in the oil and enhanced excitation efficiency due to excitation beam narrowing. While oil-layer absorption alone yields a measurable fluorescence enhancement, changes in excitation geometry significantly amplify the effect. Consequently, even small variations in  (\( t_{\text{o}} \)  and fiber-to-droplet distance can lead to substantial changes in fluorescence intensity, highlighting the importance of maintaining consistent droplet size and channel geometry in optomicrofluidic platforms.

\section{Design Perspective for Improved Fluorescence Signal}

This study identifies three key factors that enhance fluorescence signal detection from droplet-encapsulated microparticles: the particle-to-droplet size ratio (\( D^*_\text{p} \)), the particle’s radial and angular position (\( r^*_\text{p}, \theta_{\text{p}} \)), and the oil layer thickness (\( t_{\text{o}} \)) separating the droplet from the channel wall. 

First, \( D^*_\text{p} \) governs both the fluorescence intensity and its sensitivity to particle size. The fluorescence signal increases strongly from \( D^*_\text{p} \) = 0.33 to 0.5, followed by a more gradual rise between 0.5 and 0.66. This nonlinear dependence suggests two operational regimes: for size-sensitive detection, the lower-to-intermediate range (\( D^*_\text{p} \) = 0.33–0.5) provides high sensitivity to particle size. In contrast, operation at larger \( D^*_\text{p} \) values (0.5–0.66) yields higher absolute fluorescence signals with reduced sensitivity to small size variations, enabling more uniform signal levels across similarly sized particles. This ratio can be tuned by adjusting droplet diameter through flow rate control or droplet junction geometry~\cite{Lashkaripour2019,Stevens2023}. However, smaller junctions pose fabrication challenges and increase clogging risk. In platforms with shallow channels (130~ µ m), droplet spacing and stability must also be balanced.

Second, the particle position within the droplet affects both signal strength and consistency. For \( D^*_\text{p} = 0.3 \), particles near the center (\( r^*_\text{p} < 0.4 \)) yield more stable signals regardless of angle, while peripheral positions (\( r^*_\text{p} > 0.4 \)) increase angular sensitivity. Thus, placing particles near the optical axis improves reliability. Aligning detection geometry with fluid flow enhances signal capture. Pre-alignment strategies~\cite{Gaikwad2021,Yue2022}, especially guiding particles toward the channel wall adjacent to the excitation fiber (see the Supplementary Material \cite{Supplementary_Material}), improve their post-encapsulation positioning.

Third, the oil layer thickness is a critical determinant of signal strength due to its impact on optical losses. Thinner oil layers improve both excitation delivery and fluorescence collection. Simulations show nearly twofold signal enhancement when \( t_{\text{o}} \) is reduced (Fig.~\ref{FIG:7}(c)). Channel constriction near the interrogation zone~\cite{Gielen2016} or multilayer designs~\cite{Tovar2018} can help reduce \( t_{\text{o}} \), but may introduce complexity or instability. A simpler and scalable strategy is to match droplet diameter to channel height, ensuring consistent and minimal oil spacing—achievable via flow tuning and junction geometry~\cite{Lashkaripour2019,Stevens2023}.

In summary, fluorescence signal in optomicrofluidic systems can be optimized by: (i) selecting \( D^*_\text{p} = 0.5\text{–}0.66 \) for size sensitivity or \( 0.33\text{–}0.5 \) for signal uniformity, (ii) ensuring particle pre-alignment toward the excitation side for consistent post-encapsulation positioning, and (iii) minimizing oil layer thickness by matching droplet size to channel height. These experimentally accessible strategies offer a robust framework to enhance fluorescence detection performance and scalability in future droplet-based optofluidic platforms.

\section{Conclusion}
This work combined experiments and simulations to investigate light interactions with particle-encapsulated droplets in a fiber-integrated optomicrofluidic platform. We systematically studied the influence of particle-to-droplet size ratio (\( D^*_\text{p} \)), particle position (\( r^*_{\text{p}}, \theta_{\text{p}} \)), and oil layer thickness (\( t_{\text{o}} \)) on refracted and fluorescence signals. In refracted signal analysis, we identified two optical signatures: (i) droplet-refracted signal (DRS) and (ii) particle-refracted signal (PRS). DRS correlated reliably with droplet size, while PRS enabled label-free detection of encapsulated particles for \( D^*_\text{p} = 0.23\text{–}0.33 \), revealing scattering-based detection potential.

Fluorescence studies show that the signal intensity increases strongly from $D^*_\text{p}$= 0.33 to 0.5, followed by a more gradual increase up to 0.66. This defines two regimes: $D^*_\text{p}$= 0.33–0.5 provides high sensitivity to particle size, while $D^*_\text{p}$= 0.5–0.66 yields higher fluorescence intensity with reduced sensitivity to small size variations. Particle position significantly impacted signal consistency. Central positioning (\( r^*_\text{p} < 0.4 \)) minimized angular dependence, yielding uniform fluorescence, whereas off-center locations (\( r^*_\text{p} > 0.4 \)) introduced angular sensitivity and signal variability. Positioning particles closer to the optical axis is thus key for reliable detection. Reducing the oil layer thickness enhanced fluorescence by lowering optical loss at the droplet–channel interface. Matching droplet size to channel height is an effective and practical strategy for achieving stable, minimal oil layers.

Together, these findings establish a design framework for optimizing fluorescence detection in droplet-based optomicrofluidic systems. The insights presented here are applicable to high-throughput, label-free, and fluorescence-based single-cell analysis platforms, especially those requiring sensitivity to cellular heterogeneity and fine biophysical variations.

}
\vspace{4mm}

\small

{\noindent\fontsize{10pt}{16pt}\selectfont\textbf{Acknowledgments}}

The authors acknowledge support from IIT Madras via project no. RF21220988MERFIR008509, and by the Ministry of Human Resources and Development, Government of India, through the Institute of Eminence (IoE) project no. SB22231233MEETWO008509 via grant no. 11/9/2019-U.3(A).The authors would like to extend appreciation to CNNP, IIT Madras for facilitating device fabrication.

\vspace{4mm}


\printbibliography

\clearpage
\onecolumn

\begin{center}
{\fontsize{14pt}{17pt}\selectfont \textbf{Supplementary Information}}\\[1em]
\end{center}

\begin{figure}[H]
\centering
\includegraphics[width=0.7\linewidth]{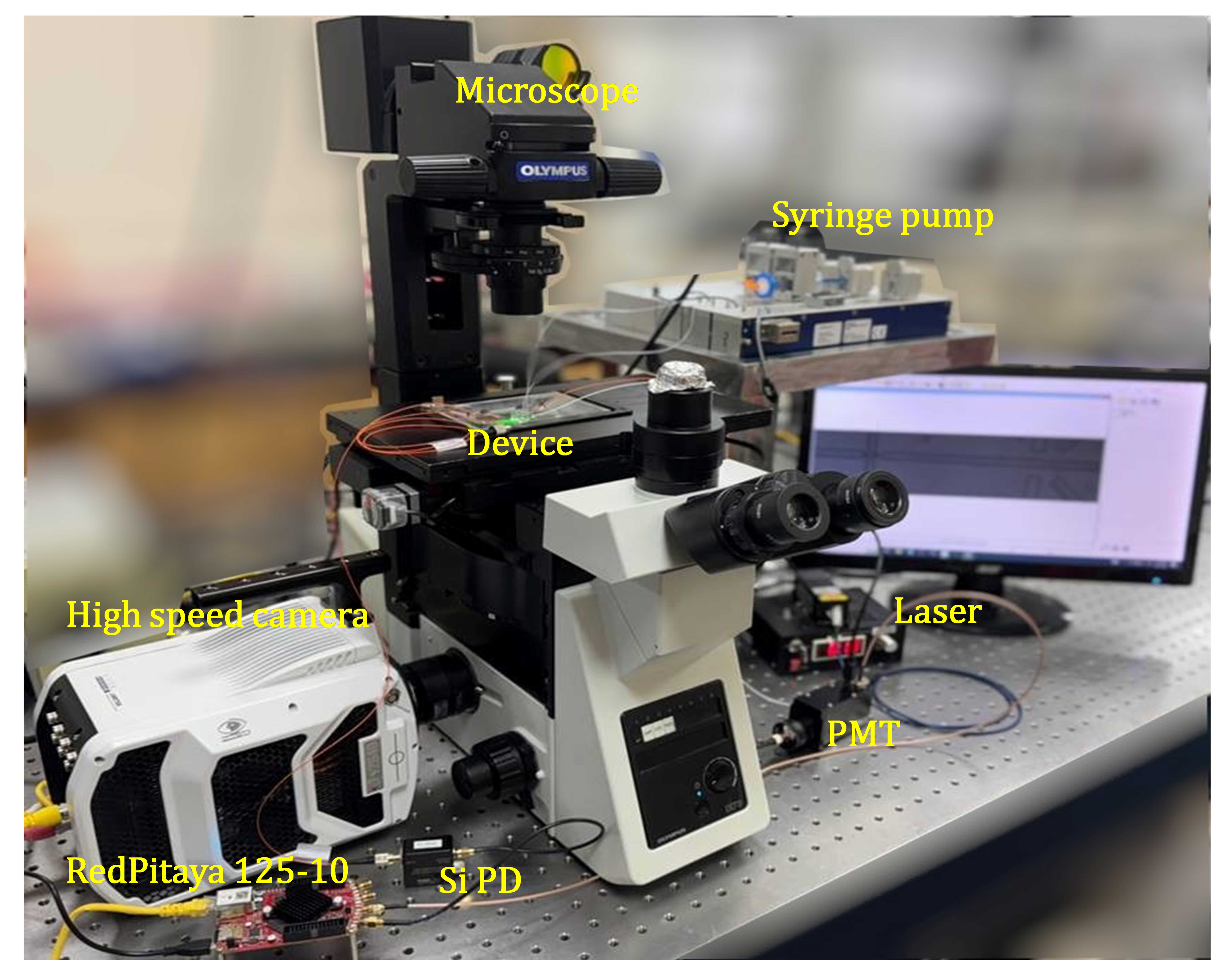}    
\caption{Photograph of experimental setup}
\label{Fig:S1}
\end{figure}

\begin{figure}[H]
\centering
\includegraphics[width=0.85\linewidth]{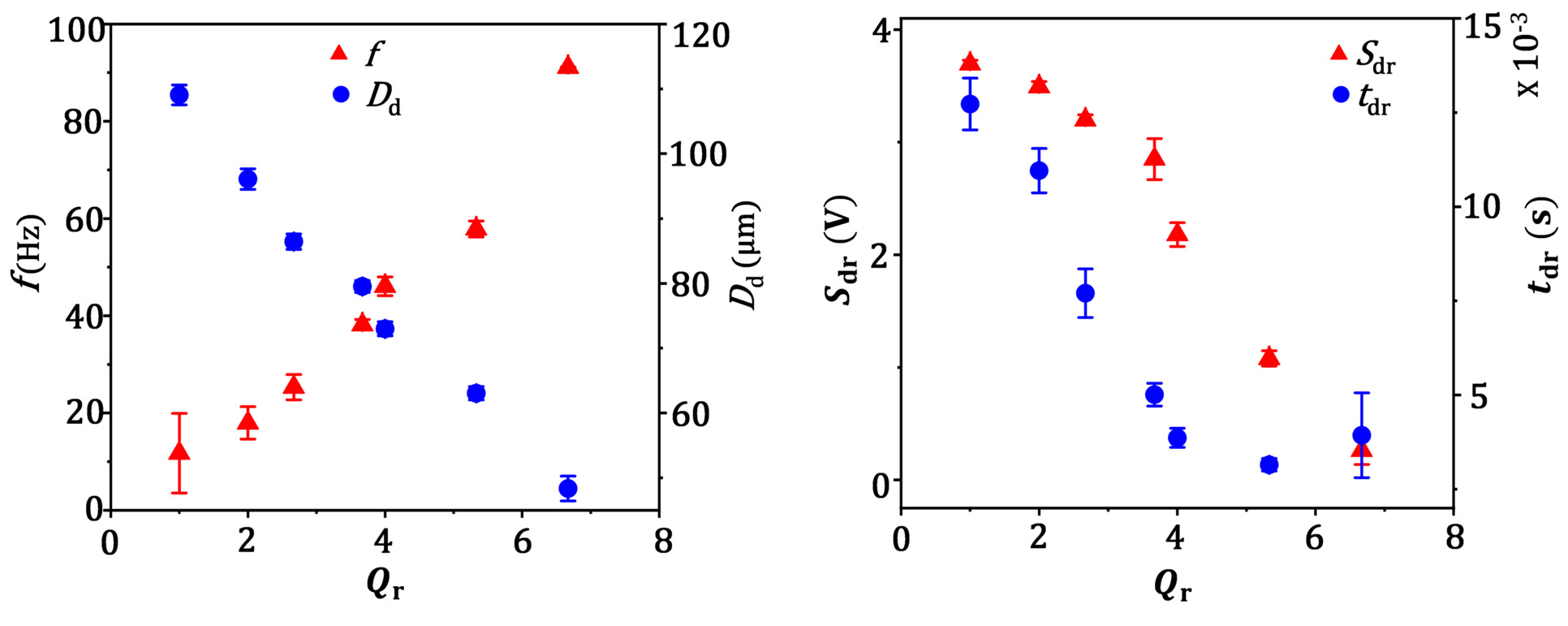}
\caption{ (a) Effect of flow rate ratio  on droplet size and droplet generation rate. (b) Variations of peak height of the DRS ($S_\text{dr}$) and peak width at 90\% peak height ($t_\text{dr}$) with flow rate ratio.}
\label{Fig:S2}
\end{figure}

\section*{\fontsize{12pt}{16pt}\selectfont Light--Droplet Interaction: Numerical Model (Ray Optics)}
This section describes the ray-optics based numerical model used to simulate light propagation and its interaction with droplets and encapsulated particles inside the microfluidic channel. The model is designed to reproduce the experimental excitation geometry and optical paths in a transparent and reproducible manner. The simulated domain corresponds to the experimentally used PDMS microfluidic device, restricted to the optical interaction region. The microchannel has a width of 100~µm, a depth of 130~µm, and a simulated length of 800~µm. The excitation single-mode fiber (SMF) has a core diameter of 9~µm and a cladding diameter of 125~µm, while the multimode fiber (MMF) has a core diameter of 62.5~µm and a cladding diameter of 125~µm. The distance between the SMF tip and the adjacent channel wall is 65~µm, and the distance between the MMF tip and the fluid channel wall is 140~µm. These dimensions match the experimental configuration. 

Light propagation is modeled using a ray-optics framework, where ray trajectories are governed by the refractive index $n$ and refraction at material interfaces is computed using Snell’s law. Each material region is treated as optically homogeneous with a constant refractive index: water ($n = 1.33$), oil ($n = 1.47$), PDMS ($n = 1.43$), single-mode fiber core ($n = 1.45$), single-mode fiber cladding ($n = 1.44$), multimode fiber cladding ($n = 1.428$), and the microparticle ($n = 1.59$). Rays propagate within each medium using the local refractive index, and the refractive index is updated at material interfaces. Ray positions are updated at fixed propagation intervals to ensure stable ray tracking along the full optical path.

To represent light emerging from the excitation SMF, rays are released from hexapolar grid points distributed across the SMF core cross-section. The ray-release plane is defined 200~µm upstream of the SMF exit facet, allowing the angular distribution of rays to develop before interaction with the surrounding PDMS and fluid domains. A hexapolar grid is used to sample the fiber core, providing uniform spatial coverage while preserving rotational symmetry. From each grid point, rays are launched within a conical distribution defined by the total internal reflection at core-clad interface of the fiber. This approach reproduces the angular divergence and spatial intensity distribution characteristic of fiber-emitted light.

\section*{\fontsize{12pt}{16pt}\selectfont Fluorescence Simulation (PDE module)}

The fluorescence model was implemented such that light diffusion was computed separately within each medium, using the corresponding diffusion and absorption coefficients for PDMS, mineral oil, and water. The excitation intensity distribution required for the fluorescence calculation was obtained from ray-optics simulations. A parametric study was performed to assess the influence of optical parameters on the simulated fluorescence signal. The parameters investigated include the absorption and reduced scattering coefficients of PDMS, mineral oil, and water, with the nominal values provided in Table~\ref{parameter}. For each parameter, the nominal value was reduced by a factor of 10 and increased by a factor of 10, while all other parameters were kept constant. The results show that variations in the absorption coefficient of PDMS have the strongest influence on the fluorescence intensity, while changes in the absorption coefficient of the oil phase have a smaller effect. Variations in the remaining parameters result in negligible changes to the fluorescence signal within the explored range.

\begin{table}[h]
\centering
\caption{Parameters for fluorescecne simulation}
\label{tab:table1}
\begin{tabular}{|c|c|}
\hline
Parameter & Value (m$^{-1}$) \\
\hline
$\mu_{a,\text{PDMS}}$  & 0.5 \cite{Lapsley2011,Greening2014}     \\
\hline
$\mu_{s,\text{PDMS}}$  & 5 \cite{Greening2014}      \\
\hline
$\mu_{a,\text{Oil}}$   & 2.3 \cite{MarnSerrano2019,Przybylek2023}    \\
\hline
$\mu_{s,\text{Oil}}$   & 0.031 \cite{AguilarArevalo2009,Gokhale2021}   \\
\hline
$\mu_{a,\text{Water}}$ & 0.2629 \cite{Buiteveld1994,Pope1997} \\
\hline
$\mu_{s,\text{Water}}$ & 0.0009 \cite{Buiteveld1994} \\
\hline
\end{tabular}
\label{parameter}
\end{table}

Parity plots corresponding to these simulations are presented in the main manuscript. The particle positions used for the parity analysis are summarized in Table~\ref{particle position}.\\

\begin{table}[h]
\centering
\caption{Particle position information used for validation of fluorescence simulation}
\label{tab:table2}
\begin{tabular}{|c|c|c|c|}
\hline
Particle number & $D_{\text{d}}$ [µm] & $\theta_{\text{p}}$  [$\degree$]& $r_{\text{p}}$ [µm]\\
\hline
P1 & 59.2  & 308  & 18 \\
\hline
P2 & 51.4 & 207 & 18 \\
\hline
P3 & 53 & 180 & 19 \\
\hline
P4 & 51.4 & 0 & 18  \\
\hline
P5 & 56.9 & 265 & 20 \\
\hline
P6 & 51.4 & 27 & 16 \\
\hline
P7 & 54.7 & 96 & 16 \\
\hline
\end{tabular}
\label{particle position}
\end{table}

\begin{figure}[H]
\centering
\includegraphics[width=0.65\linewidth]{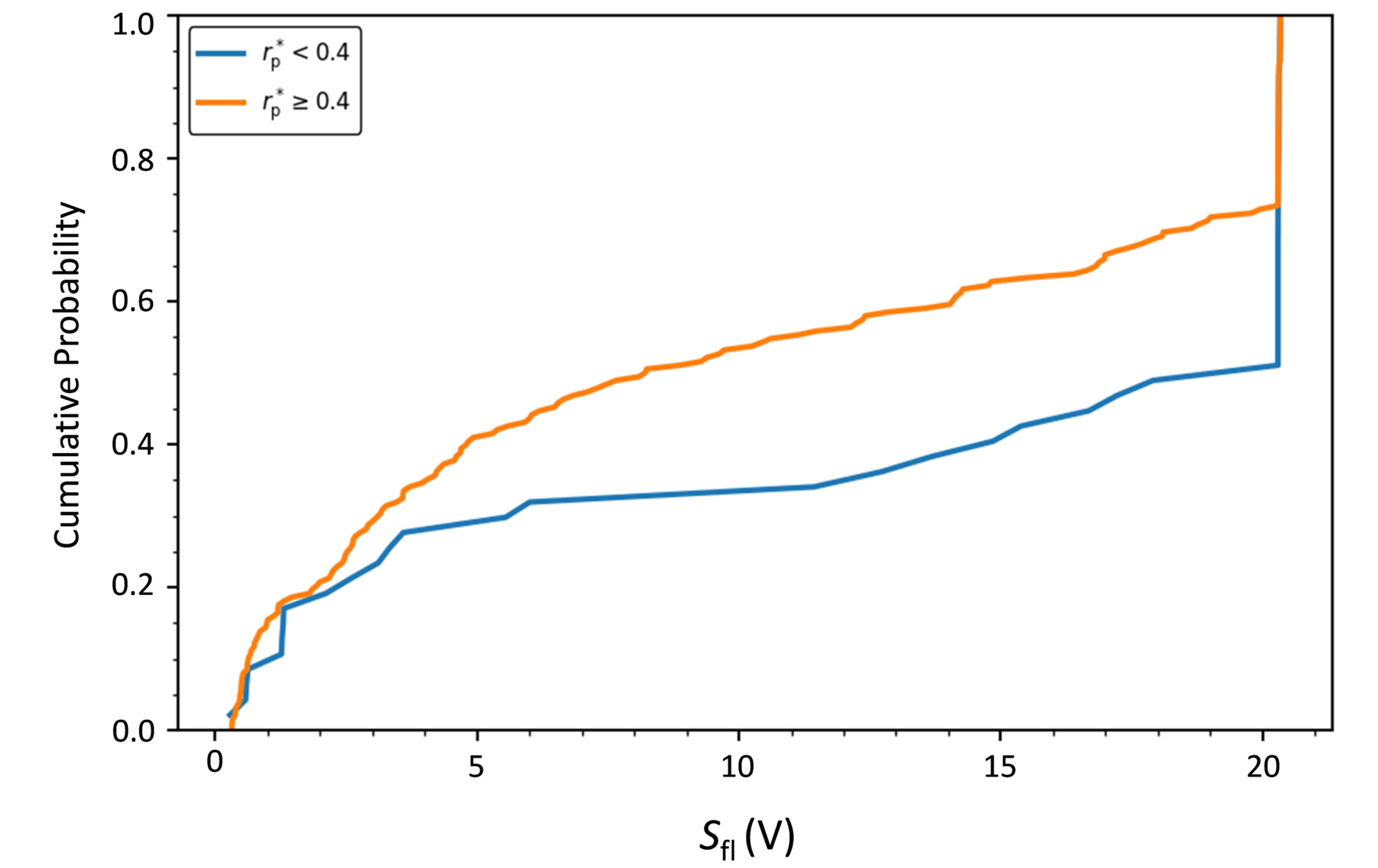}
\caption{Cumulative distribution functions of the fluorescence signal for the two $r^*_{\text{p}}$ groups.}
\label{Fig:S6}
\end{figure}

\section*{\fontsize{12pt}{16pt}\selectfont Effect of Droplet Size and Particle Position on Fluorescence Signal}

\begin{figure} [H]
\centering
\includegraphics[width=\linewidth]{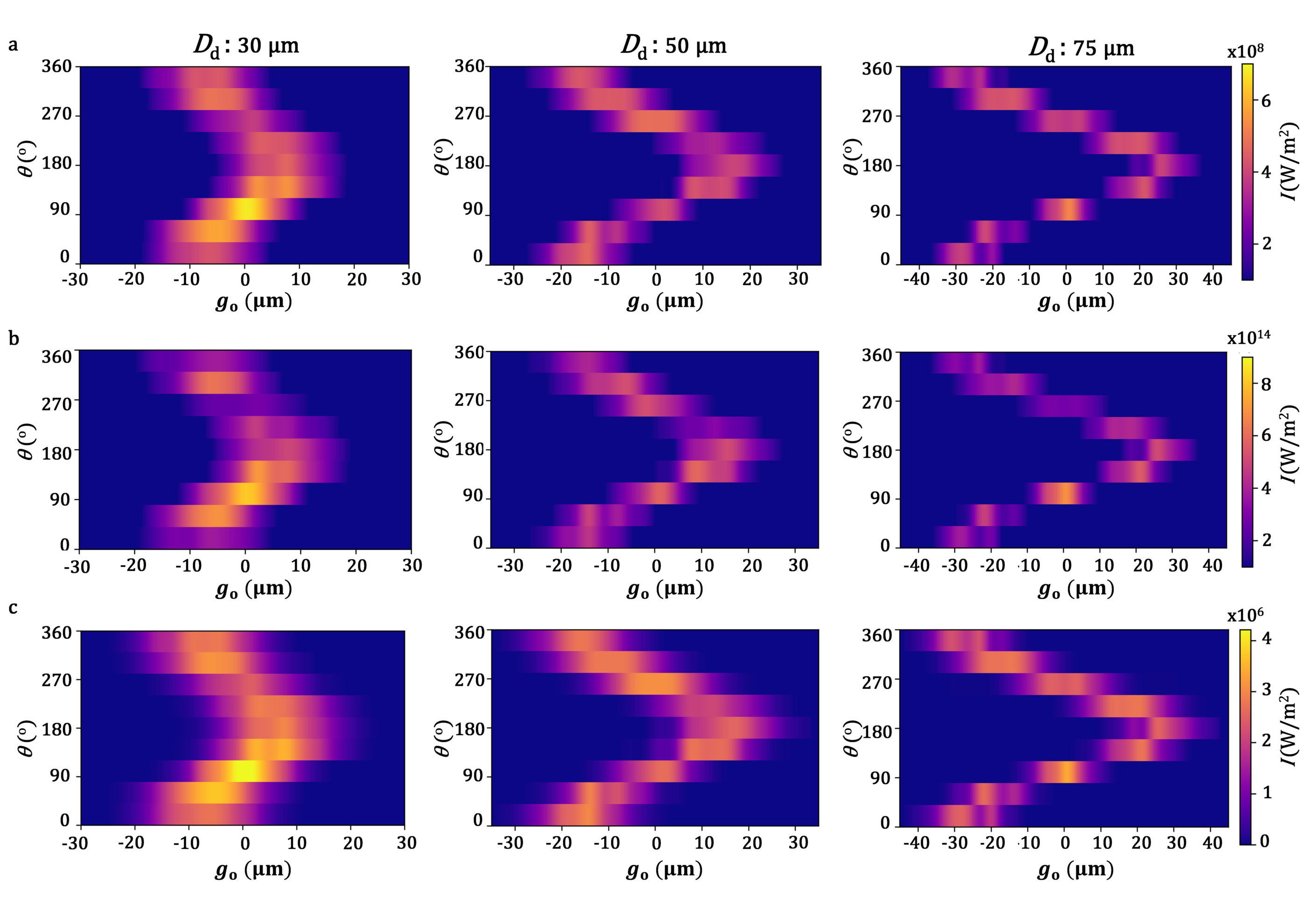}
\caption{Simulation results for varied droplet size and bead positions. (a) Light intensity reaching bead surface. (b) Bead scattering intensity reaching the 135$\degree$ MMF. (c) Fluorescence intensity from beads reaching the 135$\degree$ MMF.}
\label{Fig:S3}
\end{figure}

The impact of droplet encapsulation factors were investigated using simulations. Simulations are performed for droplets of diameters 30 µm, 50 µm, and 75 µm within a 100 µm-wide fluidic channel (\( W_{\text{c}} \)). In each case, the encapsulated particles are positioned near the droplet boundary, with a fixed $r_{\text{p}}$ such that space between particle boundary and droplet boundary is 1.5~µm and $\theta_{\text{p}}$ varied in 45$\degree$ increments. As shown in Fig. \ref{Fig:S3}, decreasing the droplet size leads to an increase in the excitation light intensity at the bead surface, as well as stronger bead scattering. The overall fluorescence signal also increases for smaller droplets. These simulation results also highlight the interplay between droplet size, oil layer thickness, and bead position on the resulting fluorescence signal intensity.

\begin{figure}[H]
\centering
\includegraphics[width=0.65\linewidth]{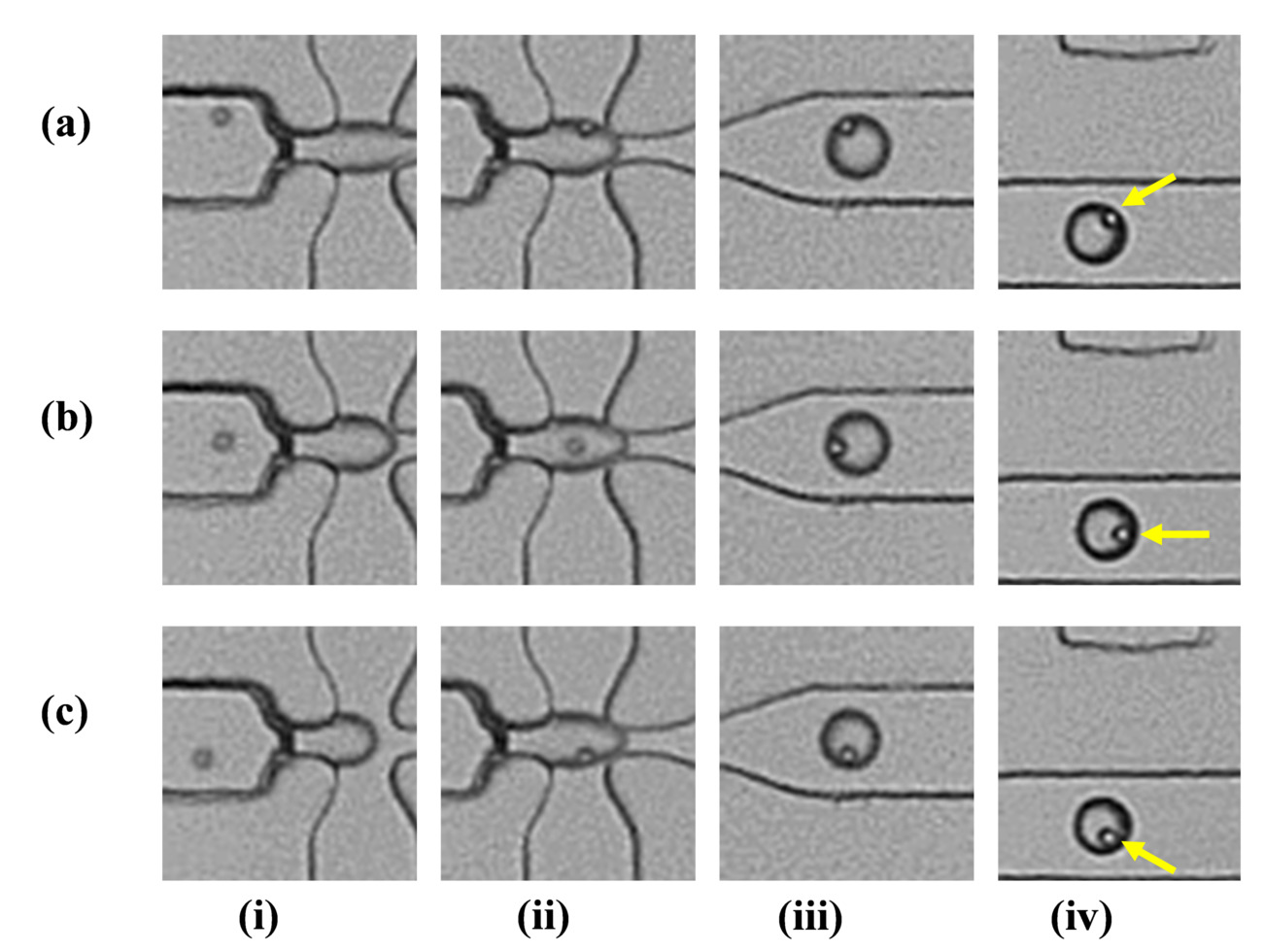}
\caption{Positional correlation between particle pre and post encapsulation. (i)-(iv) represent 4 positions of a flow focusing optofluidic device. (i) Pre-encapsulation region (ii) Encapsulation region (iii) Post-encapsulation region (iv) Light interaction region. Encapsulation dependence is shown for three regions of the droplet: (a) Region close to excitation source (b) Center region (c) Region away from the excitation source.}
\label{Fig:S4}
\end{figure}

\end{document}